\documentclass{basi}
\begin{document}

\title[Magnetic Cataclysmic Variables]{X-ray Emission from  Magnetic Cataclysmic Variables \thanks{By
            K.P. Singh, email: \texttt{Kulinderpal.Singh@tifr.res.in}}}

\author[K.~P.~Singh]%
       {K.~P.~Singh$^1$\thanks{email: \texttt{Kulinderpal.Singh@tifr.res.in}}\\
       $^1$Tata  Institute of Fundamental Research, Homi Bhabha Road,
         Mumbai 400 005, INDIA\\}

\pubyear{2013}
\volume{00}
\pagerange{\pageref{firstpage}--\pageref{lastpage}}

\date{Received \today}

\maketitle
%------------------------------------------------------------------------------%
% abstract and keywords                                                        %
%------------------------------------------------------------------------------%
\label{firstpage}
\begin{abstract}
I present a brief review of the present status of X-ray emission from Magnetic Cataclysmic Variables (MCVs).
A short introduction to the types of Cataclysmic Variables (CVs) is followed by a presentation of some of the 
properties of the two types of MCVs - Polars and Intermediate Polars (IPs) as seen in X-rays.  Finally X-ray 
spectra of MCVs and future prospects of their studies are discussed.

\end{abstract}
\section{Introduction}\label{i:Introduction}
Cataclysmic Variables (CVs) are semi-detached binary star systems with very short periods ($\sim$10 min to 10 hrs) containing
a red dwarf main-sequence-like secondary star and a more massive white dwarf (WD) primary star.
The secondary star fills its Roche Lobe and the matter overflows the lobe and is accreted by the primary white
dwarf star. CVs have been classified into various types based on their observational characteristics.  These
characteristics are in turn believed to depend on the nature of the primary star and the accretion process. 
Systems in which the white dwarf surface magnetic field is
$<$ 10$^7$ Gauss are called non-magnetic CVs and these systems display a variety of observational characteristics 
leading to further sub-types like: (a) Novae which show large eruptions of $\sim$6$-$9 magnitudes in the optical,
(b) Recurrent Novae that are previous novae seen to re-erupt, (c) Dwarf Novae (DN) that show regular outbursts of $\sim$2$-$5 
magnitudes and are further classified into SU UMa stars showing occasional super-outbursts, Z Cam stars
that show protracted standstills between outbursts, and U Gem stars that are basically all other DN, 
(d) Nova-like variables consisting of VY Scl stars which show occasional drops in brightness and UX UMa stars
consisting of all other non-eruptive variables. CVs in which the white dwarf surface magnetic field is very high, i.e.,
$\geq$ 10$^7$ Gauss, are called Magnetic CVs (MCVs). MCVs come in two varieties known as the 
Intermediate Polars (IPs) or DQ Her stars, and the Polars or AM Her stars.  The white dwarfs in the
Polars have a higher magnetic field than in the IPs. In the non-magnetic CVs the accretion takes place via 
an accretion disk produced to conserve the angular momentum of the accreting material.  
In the MCVs, however, the pressure of the magnetic field disrupts completely or partially the
accretion disk depending on the strength of the magnetic field and the accreting material is guided along the 
magnetic field lines to the poles of the white dwarf in an accretion stream or an accretion curtain
(Hameury, King \& Lasota,1986; Norton, 1993; Warner, 2003). The different types of accretion processes
are shown in Figure 1.

\begin{figure}
  \centerline{\includegraphics[angle=0,width=6cm]{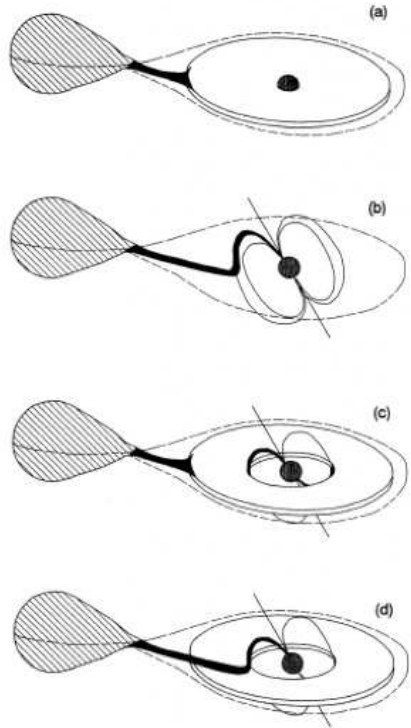}}
\caption{Schematic diagram of accretion in CVs. (a) Non-magnetic CV with no disruption of the accretion disk.
(b) Stream-fed magnetic CV in which the formation of an accretion disk is prevented. (c) Disk-fed MCV
in which the accretion disk is disrupted at the magnetospheric boundary. (d) MCV with non-accretion disk in
which the disrupted disk exists, but in which (some of) the accreting material skims over the surface to impact 
the magnetosphere directly as in (b). [Reproduced from Norton (1993)].} 
\end{figure}

Large levels of circular polarisation are seen in the optical emission from the MCVs
thus implying that the radiation process is largely cyclotron emission.
For the polarised emission to be in the optical (or UV) requires high values of 
magnetic field strength, $\geq$3x10$^7$G (3000 Tesla), which disrupts the accretion disk completely in Polars (Fig. 1(b)).
Slightly lower magnetic field leads to a partial disruption of the accretion disk seen in the IPs (Fig. 1(c) \& (d)).
The magnetically controlled accretion in MCVs is reflected in their 
eclipse light curves which show very sharp (seconds) drop and rise in the brightness, consistent with a small 
spot on the white dwarf surface. During the eclipses, different parts of the system are successively eclipsed 
and uncovered, and in some systems one can also infer the presence of accretion stream between stars that can 
be very bright, and can see faint light from the secondary star (see e.g., Bridge et al. (2002)).
As the material falling towards the WD is accreted it is de-accelerated and a shock is produced in the flow.
Gas which is initially in free-fall then encounters the shock front where the shock converts the kinetic energy 
into thermal energy (bulk motion into random motion). In this standard model of accretion onto a 
magnetic white dwarf most of the gravitational energy is released in the post-shock region. 
The resulting shock temperature is given by T$_s$ = 3GMm$_h$/8kR, which is of the order of 10 to 50 keV 
(M and R being the WD mass and radius respectively). The corresponding X- ray luminosity, L$_X$ = GMM$_{\odot}$/R, 
is $\sim$10$^{31}$ $-$ 10$^{33}$ erg s$^{-1}$ (M$_{\odot}$ being the accretion rate). Corresponding hard
X-rays are emitted mainly via the bremsstrahlung process and multi-temperature plasma are also observed 
with characteristic atomic lines from ionised plasma.   Below the shock, bremsstrahlung and cyclotron radiation 
are the two main competitive cooling processes and broad cyclotron lines are seen in the optical/UV.
Hard X-rays (E$>$2 keV) are radiated from the post-shock plasma.
The same material also radiates in IR$-$UV by cyclotron process. The accreting matter gradually settles down on the surface 
of the white dwarf. Half of the accreting luminosity illuminates the polar caps and is reprocessed as soft X-ray 
blackbody emission (E$<$2 keV) with a temperature, T$_{bb}$ = (L$_{acc}$/ 8 f $\pi$R2$\sigma$)$^{1/4}$, 
$\sim$10$-$50 eV (f being the irradiated fractional area of the stellar surface).

\section {Periods, Luminosities and Catalogues of MCVs}
In Polars, the white dwarf and the red dwarf stars are locked into the same orientation and their rotation and 
binary periods are synchronised. Consequently all the short-term variability seen in a) the orbital period radial velocity curves of 
the secondary, b) X-ray light curves from the primary, and c) optical polarisation, occurs at a single period.
The mechanism for synchronisation is believed to be the dissipation due to the magnetic field of the 
primary being dragged through the secondary.  As the relative spin rate of the primary decreases, locking can 
occur due to the dipole-dipole magneto-static interaction between primary and (weaker) secondary magnetic field. 
Polars, therefore, show only one period in their light curves or a single peak in their Fourier power spectra
if the variations are sinusoidal. Some Polars are slightly out of synchronism, and in these systems, it typically 
takes 5$-$50 days for the white dwarf orientation to repeat itself (Mouchet et al. 1999, Rana et al. 2005).  
The reason for asynchronism is not known, but one suggestion is that these systems are Polars which have 
had a recent nova event, kicking them slightly out of synchronisation (Warner 2003). 
Only four such systems are known and these can be very useful to study the effect of orientation of magnetic field 
on the accretion process.
The radiations from the IPs, however,  show a number of periods as these systems are not synchronised and the
effects of the binary rotation can be seen directly or indirectly through several peaks in the their 
Fourier power spectral densities (Norton 1993, Norton, Beardmore \& Taylor 1996, Beardmore et al. 1998, 
Rana et al. 2004, Girish \& Singh 2012 and references therein).
Wynn \& King (1992) computed theoretical power spectra of the X-ray modulations expected from IPs in various
accretion geometries, and showed that diskless accretion models show strong spin modulations and absence of
spin-orbit beat modulations.

\begin{figure}
\centerline{\includegraphics[angle=0,width=12cm]{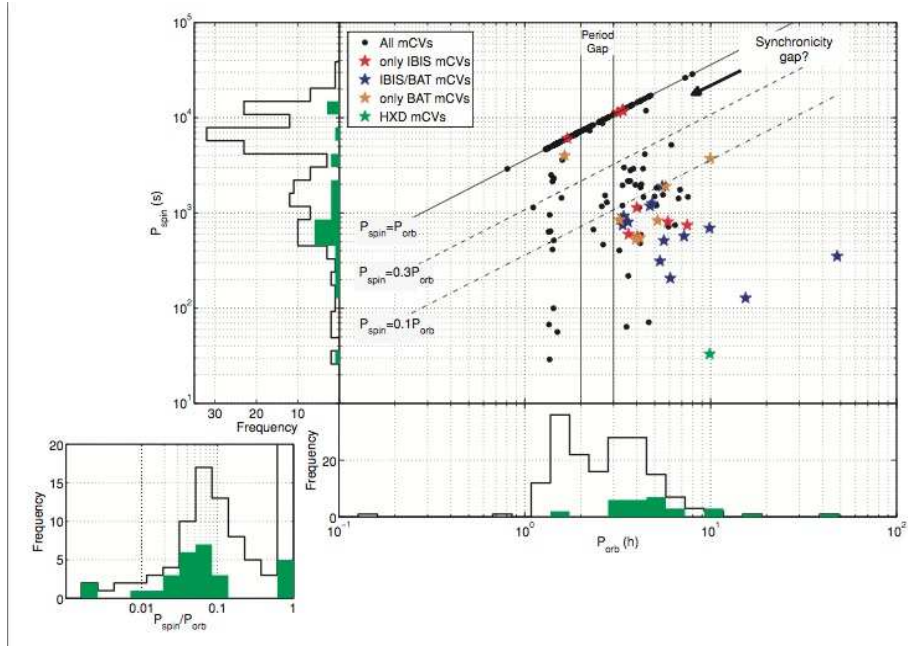}}
\caption{Orbital period versus spin period for MCVs taken from RKcat. Stars indicate MCVs detected at hard X-ray energies. Also plotted are the period gap and ÔsynchronicityÕ  lines. For reference, the orbital, spin and synchronicity distributions are also displayed. The shaded green areas represent hard X-ray-selected MCVs. [Reproduced from Scaringi et al. 2010]} 
\end{figure}

Both Polars and IPs are expected to emit high-energy photons, however, an excess of soft X-rays is observed in 
Polars (Lamb 1985). 
Chanmugam, Ray \& Singh (1991) and others have reported that the total X-ray luminosities of IPs are greater 
than those of Polars by a factor of $\sim$10, attributed mainly to the higher accretion rates. 
Moreover, it has been proposed that strong magnetic fields in Polars produce a more ÔblobbyÕ flow than in 
IPs (Warner 2003). These high-density ÔblobsÕ are then able to penetrate within the post-shock region, 
emitting fewer bremsstrahlung photons and contributing more to the soft X-ray blackbody component, 
thought to be produced at the base of the post-shock region.
In recent years, an increasing number of IPs have been detected and discovered by hard X-ray telescopes 
such as INTEGRAL/IBIS (Barlow et al. 2006; Landi et al. 2009), Swift/BAT (Brunschweiger et al. 2009) 
and SUZAKU/HXD (Terada et al. 2008). Scaringi et al. (2010) have compiled a new list of IBIS MCVs, forming 
part of a larger compilation of hard X-ray-selected IPs from all three telescopes mentioned above, 
to help study the general properties of this hard X-ray-selected sample.  Scaringi et al. (2010) used 
two well known Catalogues of the CVs: 1) the Catalogue and Atlas of Cataclysmic Variables 
(Downes et al. 2005, or DWScat) and the Catalogue of Cataclysmic Binaries (Ritter \& Kolb 2003, or RKcat). 
DWScat contains 1830 CVs, whereas RKcat contains 731. Scaringi et al (2010),  merged both the catalogues and found 
that 656 CVs are common to both the catalogs. 
The principal reason for RKcat having fewer objects is that only CVs with known orbital periods are included in the sample; 
however, RKcat also includes a few CVs that DWScat does not report (Scaringi et al 2010). Their final known CV set therefore 
contains 1905 CVs (hereafter DRKcat). Fewer than $\simeq$ 9\% of the total number of CVs within DRKcat are known to be 
magnetic in nature (Polars and IPs), and only $\simeq$3\% (56 sources) are IPs.
Compared to IPs, only 4 Polars are detected at high energies with INTEGRAL, including the two asynchronous 
systems BY Cam and V1432 Aql (Landi et al. 2009, Scaringi et al. 2010). 
Thus, though optically selected samples favour Polars, with the number of known Polars being twice that of IPs, 
in the hard X-ray-selected sample, $\sim$90\% of the MCVs detected are IPs. 
This is not unexpected since the IPs are seen to produce 10 times more hard X-rays than Polars due to their higher mass transfer 
and intrinsically harder spectrum (see Warner 2003).
The distribution of the orbital and spin periods in this expanded sample of MCVs is shown in Figure 2 
reproduced from Scaringi et al. (2010). They find that most hard X-ray-detected MCVs have P$_{spin}$ /P$_{orb}$ $< $0.1 
above the period gap, and that there are very low number of detected systems in any band between P$_{spin}$ /P$_{orb}$ = 0.3 
and P$_{spin}$ /P$_{orb}$ = 1 and the apparent peak of the P$_{spin}$ /P$_{orb}$ distribution at about 0.1.

\section { X-ray Spectra of MCVs}
Broad-band (IR to UV) spectrum observed from a Polar - AM Her is shown in Figure 3
taken from Beuermann (1999). Apart from the photospheric emission from the white dwarf seen in UV,
emission seen in the IR-Optical-UV bands is from cyclotron radiation from accretion column. The dominating
soft X-ray emission is seen as a blackbody emission from the heated surface of the WD in Polars. 
The absence of the blackbody emission in the IPs, in the early observations, 
was interpreted as due to either strong absorption in the accretion curtain or to a lower temperature EUV 
emission from a larger irradiated zone. 
The number of IPs detected in soft X-rays (E$<2$keV) has recently increased significantly 
(Evans \& Hellier 2007, Anzolin et al. 2008, Girish \& Singh 2012). The black-body soft-X-ray component in the IPs
has temperatures ($\sim$50-120 eV) that are higher than those usually found in Polars (20-60eV) and 
with a lower soft/hard flux ratio. The higher temperatures found in IPs indicate higher accretion rates, 
in line with higher luminosities.
Multiple and partial absorption is also seen in the soft X-ray spectra of IPs (e.g., see Girish \& Singh 2012) indicating
the presence of complicated patterns of absorption by accretion curtains in the line of sight.
The maximum temperature of the system seen in the hard X-ray continuum provides an estimate of 
the mass of the accreting magnetic white dwarf (see \S1).

A prominent emission line, presumably from ionised Fe in hot plasma can be seen in almost all CVs. 
The Fe line complex is resolved in very high energy resolution spectra taken with Chandra High 
Energy Transmission Gratings (HETG), and an example is shown in Figure 4 taken from Girish, Rana \& Singh (2007).
The intensity of the He-like triplets seen at 6.7002 keV (r: resonance line), two inter combination lines at 6.6821 keV 
and 6.6673 keV (i=i1+i2) (unresolved), and 6.6364 keV (f: forbidden) line provide a powerful diagnostics of the ionised plasma 
in the hot or highly ionised plasma.  For example, assuming a thin plasma, as in the solar corona, 
G-ratio = (f+i)/r can be used to get the electron temperature, and R-ratio = f/i to obtain the electron density of the plasma,
Similarly, H-like doublets at 6.973 keV  and 6.952 keV indicate the highest temperature regions.
The X-ray spectrum actually contains H-like and He-like lines of many other elements as well:
S, Si, Mg, Ne and O with several Fe L-shell emission lines. The forbidden lines in the spectrum are generally 
weak whereas the H-like lines are stronger suggesting that emission from a multi-temperature, collissionally 
ionised plasma dominates in AM~Her.  A line-by-line fitting analysis of all the available spectra on all CVs obtained using 
the Chandra HETG has been reported by Schlegel et al. (2013),  who report the existence of broad ionisation 
and electron temperatures ranging from $\sim$0.4 keV to $\sim$5$-$10 keV, thus confirming the previous results 
based on a global analysis of HETG data by Hellier \& Mukai (2004). 
Number densities are also found to cover a broad range, from 10$^{12}$ to $>$ 
10$^{16}$ cm$^{-3}$ (Singh et al. 2006, Rana et al. 2006, Schlegel et al. 2013). 
Much of the plasma is probably in a non-equilibrium state; the Fe emission, however, may 
arise from plasma in ionisation equilibrium (Schlegel et al. 2013).

\begin{figure}
\centerline{\includegraphics[angle=0,width=8cm]{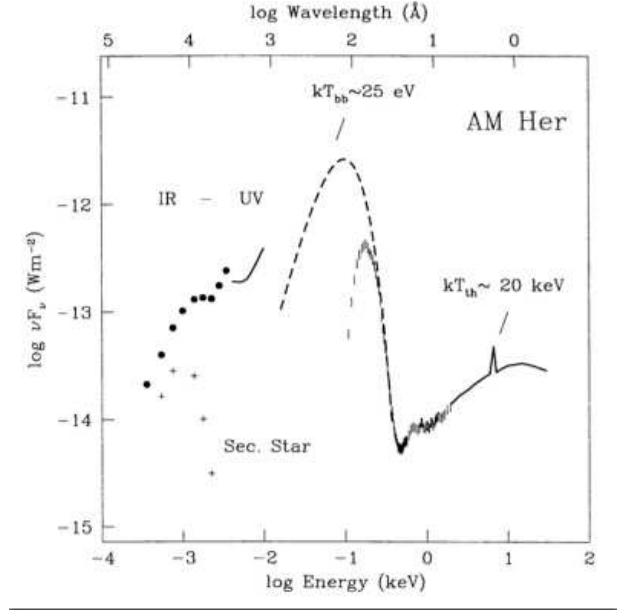}}
\caption{Multi-wavelength spectral energy distribution of the Polar AM Her [Reproduced from K. Beuermann, 1999].} 
\end{figure}

\begin{figure}
\centerline{\includegraphics[angle=0,width=12cm]{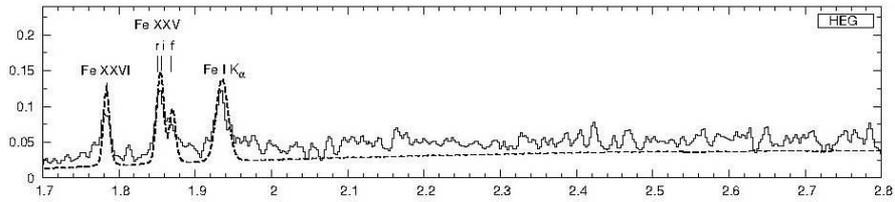}}
\caption{Chandra HETG spectrum of AM Her. A section of the spectrum near the Fe line complex is shown. [Reproduced from Girish, Rana \& Singh (2007)]} 
\end{figure}

Phase resolved spectroscopy of AM Her has shown that the line centres of 
%\ion{Mg}{12},
Mg {\rm XII},
%\ion{S}{16}, 
S {\rm XVI}, resonance line of 
%\ion{Fe}{25}, 
Fe {\rm XXV}, and 
%\ion{Fe}{26} 
Fe {\rm XXVI} emission are modulated by a few hundred to 1000 km s$^{-1}$ 
from the theoretically expected values, indicating bulk motion of ionised matter in the accretion column of 
AM~Her (Girish, Rana \& Singh 2007).  The observed velocities of Fe {\rm XXVI} ions are close to the expected shock 
velocity for a $\sim$1M$_\odot$ white dwarf.  The observed velocity modulation is consistent with that expected from a single
pole accreting binary system (Girish et al. 2007).
Almost all CVs (including the MCVs) show most of these lines from ionised species, and also show a line due to fluorescence of Fe 
(Fe {\rm I} K$_{\alpha}$) at 6.4 keV, thus indicating the presence of cold (un-ionised) material and reflection in both 
the disk and diskless systems (Ezuka \& Ishida 1999, Rana et al. 2006; Girish et al. 2007; Girish \& Singh 2012).

\section {Future}
Simultaneous multi-wavelength studies in optical, UV, soft and hard X-rays with ASTROSAT of the newly discovered 
population of IPs (INTEGRAL sources), would be of great help in elucidating the nature of these sources.
Broad-band hard X-rays continuum observations of MCVs with ASTROSAT will provide better estimates 
of the mass of the accreting WDs.
High spectral resolution studies with X-ray calorimeter onboard  Astro-H will provide the ionisation structure of the accretion 
column and the properties of the hot multi-temperature plasma behind the shock front in the accretion column.

\end{document}